\documentclass[twocolumn,showpacs,aps,prl,groupedaddress]{revtex4-1}
\usepackage{epsfig}
\usepackage{amsmath}
\usepackage[switch, modulo]{lineno}

\include{def}
\def\Journal#1#2#3#4{{#1} {#2} (#3) #4}

\def\JPCS{J.Phys.Conf.Series}
\def\JPG{J.Phys.G}
\def\EPJ{Eur.Phys.J.}
\def\ARNPS{Ann.Rev.Nucl.Part.Sci.}

\def\PLB{Phys.Lett.B}

\def\PRL{Phys.Rev.Lett.}
\def\PRC{Phys.Rev.C}

\def\PR{Phys.Rev.}

\def\EPJ{Eur.Phys.J.}
\def\PAN{Phys.Atom.Nucl.}

\def\APP{Acta Phys. Pol.}
\def\be{\begin{equation}}
\def\ee{\end{equation}}

\begin{document}

\title{Core-corona interplay in Pb-Pb collisions at $\sqrt{s_{NN}}$=2.76 TeV}

\author{M.Petrovici, I.Berceanu, A.Pop, M.T{\^ a}rzil{\u a}, C.Andrei}

\affiliation{National Institute for Physics and Nuclear Engineering - IFIN-HH\\
Hadron Physics Department\\
Bucharest - Romania}

\date{\today}

\begin{abstract}

    A simple approach based on the separation of wounded nucleons in an A-A collision 
 in two categories, those suffering single collisions - corona and the rest - core, estimated within 
 a Glauber Monte-Carlo model, 
 explains the centrality dependence of the light flavor hadrons production in Pb-Pb collisions at 
 $\sqrt{s_{NN}}$=2.76 TeV. The core contribution
 does not include any dependence of any process on the fireball shape as a function of the impact parameter.
 Therefore, the ratios of the $p_T$ distributions to the one corresponding to the minimum bias pp collisions at the
 same energy, each of them normalised to the corresponding charged particle density, the $\langle p_T\rangle$ and transverse energy per unit of rapidity  
 are reproduced less accurate by such an approach. The results show that at LHC energies the corona contribution still plays an important role and it has to be considered in order to evidence the centrality dependence of different observables related to the 
 core properties and dynamics.  
    
\end{abstract}


\maketitle

\section{}

\subsection{}
\subsubsection{}

\vspace{-3.1cm}

   At ultra-relativistic energies the initial energy density distribution within the fireball is highly non-homogeneous
\cite{Dre}. This non-homogeneity is directly correlated with the distribution 
of nucleons suffering 
a given number of collisions (the thickness of nuclear matter to which a nucleon is exposed) 
which can be estimated in a simple Monte-Carlo Glauber approach. 
The non-homogeneous distribution of the initial state has a direct consequence on the fluctuations of different observables and 
the dynamics 
of the initial state up to the freeze-out moment. At a given collision energy, the amount of energy transferred into the fireball 
increases with centrality, i.e. the volume of the overlapping zone, and is correlated with the measured charged particle multiplicity.
Therefore, the charged particle multiplicity dependence was extensively studied for many observables.
It is mandatory to understand how much of this dependence on centrality or system size is due to 
the initial high energy density state and its subsequent dynamics and how much is due to a simple interplay between
core-corona contributions.
   
  Extensive studies of the core-corona separation effects on the centrality dependence of the strangeness production, average transverse momenta, 
elliptic flow or $p_T$ spectra in heavy-ion collisions at SPS and
RHIC energies were done \cite{Bec0, Boz, Wer, Stei, Bec, Aich, Boz1, Gem, Schr}.
The core-corona interplay estimated by 
parametrizing the core-mantle contributions such as to reproduce the density of charged particles produced in Au-Au collisions 
\cite{Boz} was confirmed by more consistent approaches,
based on string density before hadronization within the EPOS model using some global parameters \cite{Wer} or on quark number density within the UrQMD transport+hydrodynamics hybrid model \cite{Stei}.
Approaches based on separating the participating (wounded) nucleons $N_{part}$, in two classes, those which scatter only once, coined corona, 
which would correspond to a minimum-bias pp collision and the rest which contribute to the fireball, called core, both estimated within the
geometrical Glauber model,  were equally successful in explaining the experimental trends \cite{Bec, Aich}.
A very good agreement between the percentage of nucleons undergoing single collisions
and the percentage taken as a fit parameter in a core-corona scenario  was evidenced \cite{Boz1}.        
  Based on this and due to its simplicity, in the present paper we consider the approach from \cite{Aich} in order to explore to which extent it 
can explain the centrality dependence of  different observables in Pb-Pb collisions at $\sqrt{s_{NN}}$=2.76 TeV.  

The number of participating  nucleons and the 
percentage of wounded nucleons which scatter more than once, $f_{core}$, were estimated using the Monte-Carlo 
Glauber approach \cite{Gla, Fra, San}. For the nuclear density profile of the colliding Pb nuclei  a Woods-Saxon distribution was used:   
\begin{equation}
\rho(r)=\frac{1}{1+exp(\frac{r-R_A}{a})}
\end{equation}    
\noindent
with a=0.546 fm and 
$R_A$=6.62 fm.
The nucleons of the two Pb nuclei are colliding if the transverse distance between them is smaller than the 
distance estimated from the inelastic nucleon-nucleon cross section of 64 mb at $\sqrt{s}$=2.76 TeV \cite{Alcen, Al5}. 
   The correlation between the centrality and 
impact parameter is taken from \cite{Alcen}.
\begin{figure}[t!]
\centering{\mbox{\epsfig{file=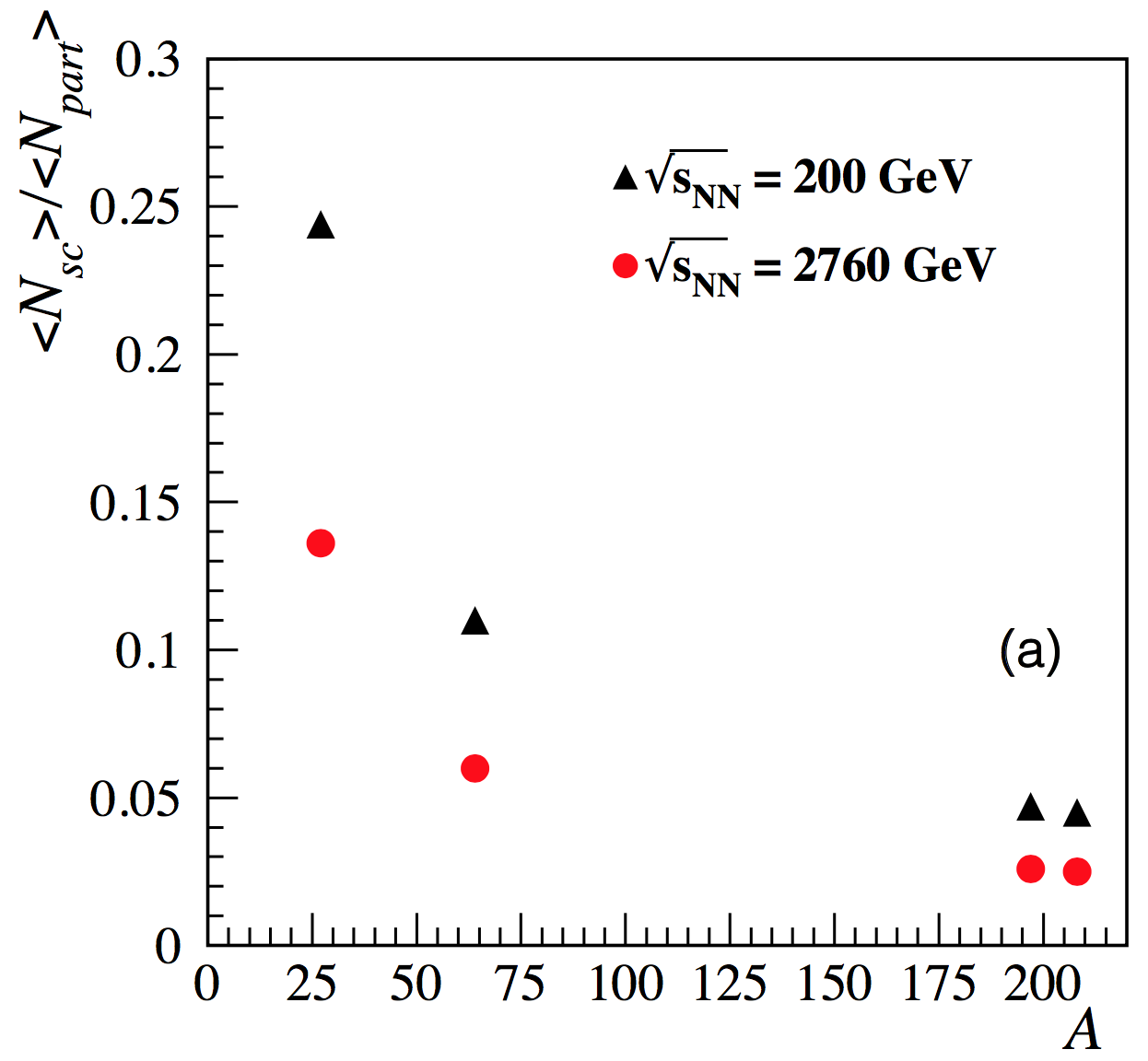, width=.45\textwidth}}}
\centering{\mbox{\epsfig{file=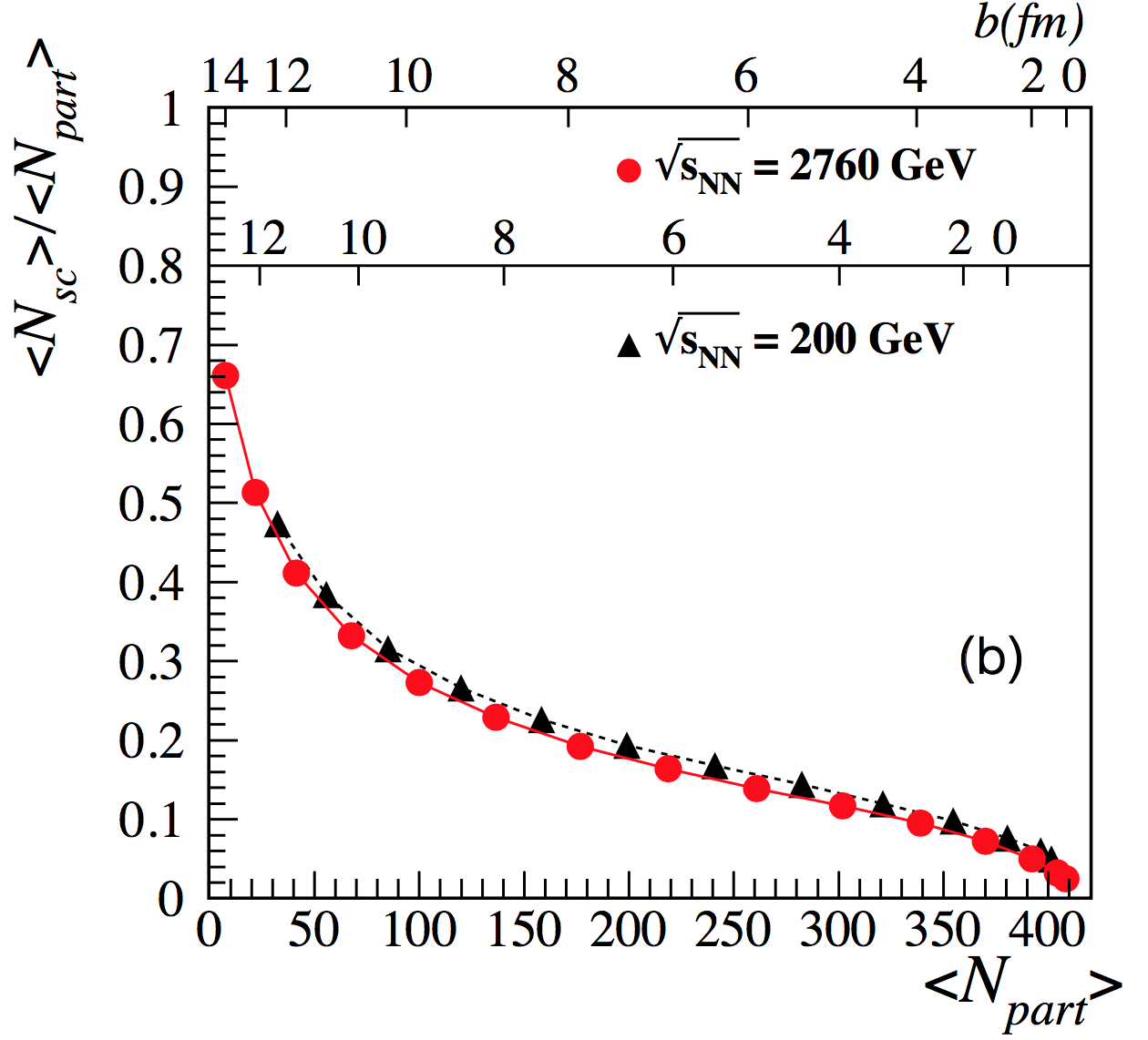, width=.475\textwidth}}}
\caption{a) The percentage of wounded nucleons which scatter only once,  for head-on
collisions of two identical nuclei, as a function 
of their mass at $\sqrt{s_{NN}}$=200 GeV and  2.76 TeV; \\
b) The same percentage as a function of $\langle N_{part}\rangle$
in Au-Au at $\sqrt{s_{NN}}$=200 GeV and Pb-Pb at $\sqrt{s_{NN}}$=2.76 TeV collisions.}
\label{fig-1}
\end{figure}
These values, $\langle N_{part}\rangle$ and the percentage of nucleons suffering more than a single collision are presented in Table I.
\begin{table}[h]
\begin{center}
\begin{tabular}{cccccc}
 $Cen$ (\%)    & b (fm) & & $\langle N_{part}\rangle$ & & $f_{core}$ \\
\hline
0 - 5     &   0.00 -  3.50  & & 382.5$\pm3.1$  & & 0.942$\pm 0.003$   \\
5 - 10   &   3.50 -  4.95  & & 329.4$\pm4.9$  & & 0.900$\pm 0.002$   \\
10 - 20   &   4.95 -  6.98  & & 259.9$\pm2.9$  & & 0.861$\pm 0.002$   \\
20 - 30   &   6.98 -  8.55  & & 185.4$\pm3.9$  & & 0.814$\pm 0.002$  \\
30 - 40   &   8.55 -  9.88  & & 128.1$\pm3.3$  & & 0.764$\pm 0.002$  \\
40 - 50   &   9.88 - 11.04 & &   84.2$\pm2.6$  & & 0.703$\pm 0.002$ \\
50 - 60   & 11.04 - 12.09 & &   52.1$\pm2.0$  & & 0.626$\pm 0.002$  \\
60 - 70   & 12.09 - 13.05 & &   29.5$\pm1.3$  & & 0.536$\pm 0.004$ \\
70 - 80   & 13.05 - 13.97 & &   14.9$\pm0.6$  & & 0.431$\pm 0.005$  \\
80 - 90   & 13.97 - 14.96 & &    6.3$\pm0.2$  & & 0.325$\pm 0.003$ \\
\hline
\end{tabular}
\caption{Centrality, impact parameter \cite{Alcen} $\langle N_{part}\rangle$ and the percentage of wounded nucleons with more than one 
collision estimated for the corresponding centrality, based on the Glauber Monte-Carlo approach, 
for Pb-Pb at
$\sqrt{s_{NN}}$=2.76~TeV collisions.}
\label{tab:table2}
\end{center}
\end{table}
\noindent
   In Fig.1a the percentage of wounded nucleons which scatter only once,  for head-on
collisions of two identical nuclei is represented as a function 
of their mass at two incident energies, i.e. $\sqrt{s_{NN}}$=200 GeV and 2.76~TeV. 
The percentage of nucleons scattered only once is decreasing 
with the increasing mass of the colliding nuclei and the centre of mass energy, the value for Pb-Pb collisions at the LHC energy being very small.
However, for peripheral and mid-central collisions their contribution becomes important and a good scaling is evidenced as a function 
of the number of participating nucleons for Au-Au and 
\mbox{Pb-Pb} collisions at the respective energies, as could be followed in Fig.1b.  

We start with integrated values,
i.e. the yields of light flavor hadrons per unit of rapidity measured in the mid-rapidity region \cite{Al3,Al1,Al2,Al14}.
Within a core-corona approach, the yield of a given species $i$, at a given centrality, per unit of rapidity can be written as:
\begin{equation}
\left( \frac{dN}{dy}\right) _i^{cen}=N_{part}[(1-f_{core})M_i^{pp MB} + f_{core}M_i^{core}]
\end{equation}
\noindent
where $M_i^{ppMB}$=$\frac{1}{2}(dN/dy)_i^{ppMB}$ at the same energy and 
$M_i^{core}$ is the multiplicity per core
participant.  A similar decomposition can be done for the double differential $p_T$ distributions.
$(dN/dy)_i^{ppMB}$  values for $\mathrm{\pi}$, K, p were reported in \cite{Al13}.
The values for hyperons and $\phi$ mesons were obtained following the same receipt as the one used by the ALICE Collaboration \cite{Al7,Al10,Al2,Al14,Al16} and reported values \cite{Al20}, respectively. 
Eq.(2) for \mbox{0-5$\%$} centrality was
used to extract the core contribution for each species.
\begin{figure}[h]
\centering{\mbox{\epsfig{file=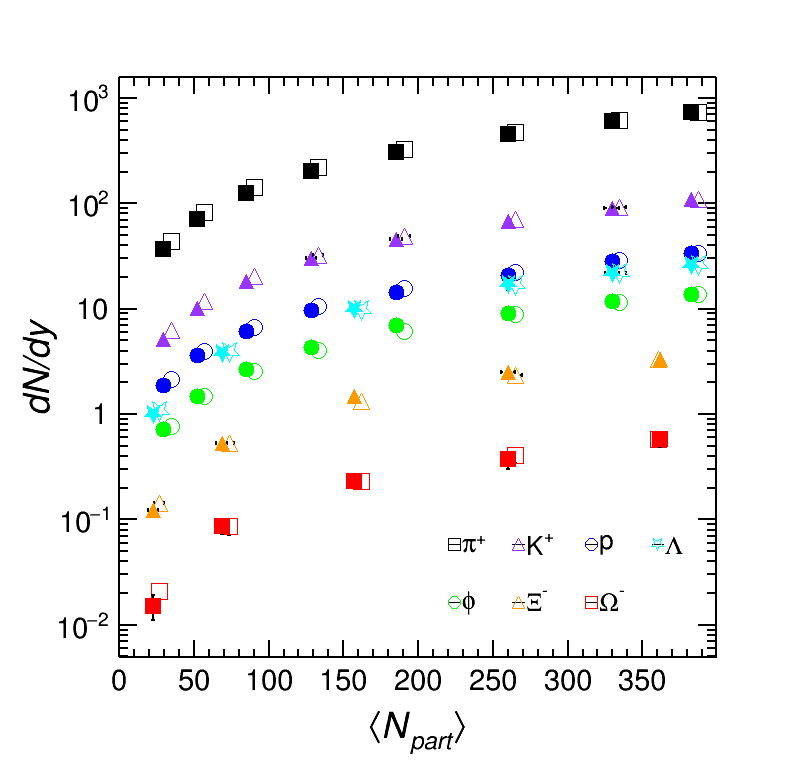, width=.47\textwidth}}}
\vspace{-.5cm}
\caption{The estimated light flavor hadrons yields as a function of 
participants $\langle N_{part}\rangle$ for Pb-Pb collision at $\sqrt{s_{NN}}$=2.76~TeV:
using Eq.(2) - open symbols;
experimental values - full symbols \cite{Al3,Al1,Al2,Al14}. To be visible, the open symbols are shifted by 5 units in $\langle N_{part}\rangle$.}
\label{fig-2}
\end{figure}
\begin{figure*}[ht]
\centering{\mbox{\epsfig{file=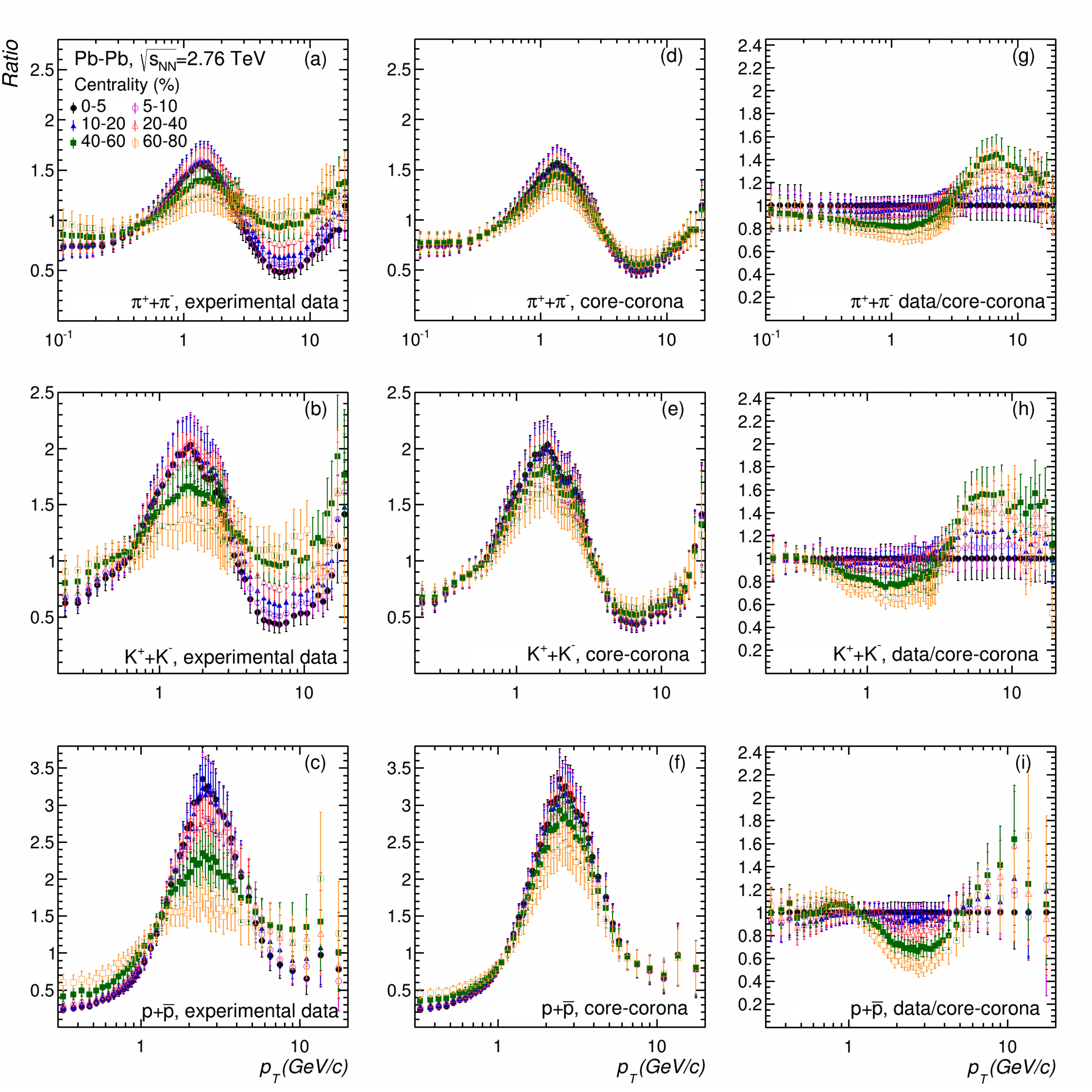, width=0.88\textwidth}}}
\caption{The centrality dependent ratios of normalised $p_T$ spectra, measured in Pb-Pb collisions at $\sqrt{s_{NN}}$=2.76 TeV,
to the normalised minimum-bias spectrum at the 
same energy. The spectra are normalised to the corresponding average charged particle density. Left column (a, b, c): based on $p_T$ spectra measured by the ALICE Collaboration
\cite{Al9, Al11}; Middle column (d, e, f): the results using the core-corona scenario; Right column (g, h, i): the ratios of the experimental $p_T$ distributions relative to the ones estimated within the core-corona approach. From top to
bottom they correspond to pions, kaons and protons.}
\label{fig-3}
\end{figure*}
\noindent
The estimated light flavor hadrons yields as a function of the average number of participants $\langle N_{part}\rangle$ from Eq.(2)
and the published experimental values are presented in Fig.2 by open and full symbols, respectively. 
As it is 
observed, the enhancement of light 
flavor hadrons production with 
increasing centrality is reproduced rather well by a pure geometrical approach, a superposition of the single 
nucleon-nucleon (corona) contribution and the contribution due to nucleons which scatter more than once (core) whose
relative weights were estimated within the Glauber approach. 
This approach is also applied for the ratio of double differential cross sections normalised to the
charged particle densities, i.e.: 
$\left(\frac{\frac{d^2N}{dydp_T}}{\langle \frac{dN_{ch}}{d\eta}\rangle }\right)_i^{cen}/\left(\frac{\frac{d^2N}{dydp_T}}{\langle \frac{dN_{ch}}{d\eta}\rangle }\right)_i^{ppMB}$. 
The numerator and denominator of $\left(\frac{\frac{d^2N}{dydp_T}}{\langle \frac{dN_{ch}}{d\eta}\rangle }\right)_i^{cen}$ were
estimated separately as a superposition of corona and core contributions.
Similar to the integrated yield, the core contributions are obtained from the experimental spectra for the most central collisions. 
In Fig.3 left column (a,b,c), the ratios of the normalised $p_T$ distributions \cite{Pet2} based on $p_T$ spectra 
measured by the ALICE Collaboration \cite{Al9, Al11} are represented. 
The results using the core-corona scenario can be followed in the second column (d,e,f) while the ratios 
between experimental $p_T$ distributions and the ones estimated within the core-corona approach are shown in the third column (g,h,i). 
The trends observed in the experiment 
are qualitatively reproduced.
Quantitatively, as could be seen in the last column of Fig.3, there are some differences between the estimates based on the core-corona ansatz and the 
experimental values. 
This is the consequence of not including
any dependence of different phenomena on the core shape as a function of centrality. At RHIC energies, the PHENIX Collaboration has 
done detailed studies on the azimuthal dependence of the in-plane/out-of-plane transverse expansion \cite{Phe2}  and $R_{AA}$ \cite{Phe1}
for Au-Au collisions at $\sqrt{s_{NN}}$=200 GeV. 
A clear $N_{part}$ and azimuthal angle dependence of these two phenomena was evidenced. 
This is the explanation of the observed deviations in the last column of Fig.3 in the region of $p_T$ where the expansion and suppression have the 
main contribution to the spectra shapes.  
Therefore, a deviation from a simple core-corona 
approach where the core contribution scales only with the number of participants colliding 
more than once, not including any dependence on the 
fireball shape, is expected. 
Being aware of these effects, one can investigate to which extent the $\langle p_T\rangle $ dependence as a function of centrality is predicted 
in the simple core-corona approach by the formula:
\begin{multline}
\langle p_T\rangle _i^{cen}=\\
\frac{f_{core}\langle p_T\rangle_i^{core}M_i^{core}+(1-f_{core})\langle p_T\rangle_i^{ppMB}
M_i^{ppMB}}
{f_{core}M_i^{core}+(1-f_{core})
M_i^{ppMB}}
\end{multline}
\noindent
$\langle p_T\rangle _i^{pp MB}$ for $\mathrm{\pi}$, K, p in pp collisions at 2.76 TeV were reported by the ALICE 
Collaboration \cite{Al13}. 
\begin{figure}[h]
\vspace{-0.4cm}
\includegraphics[width=0.47\textwidth]{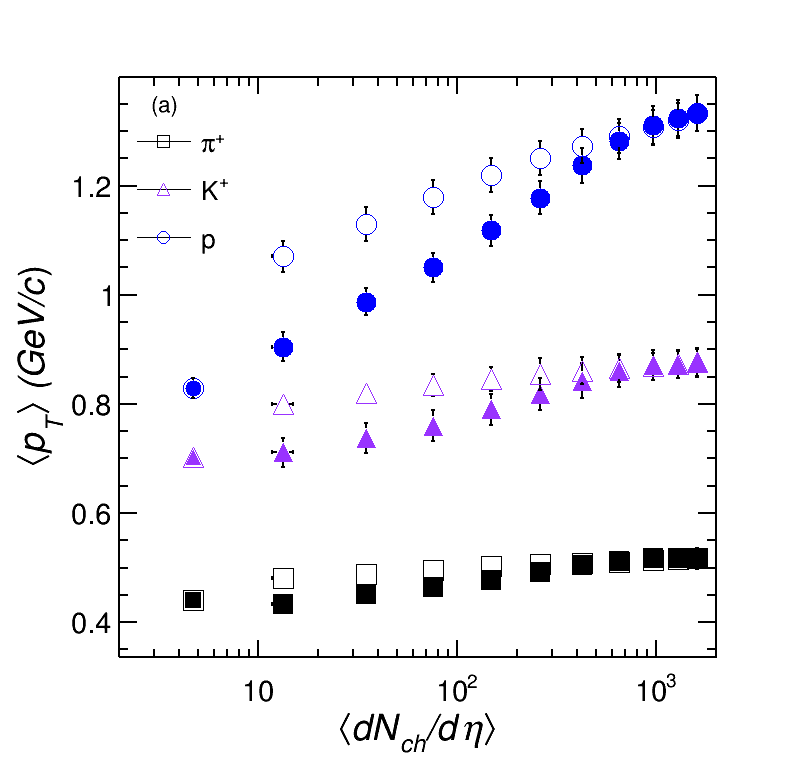}
\includegraphics[width=0.47\textwidth]{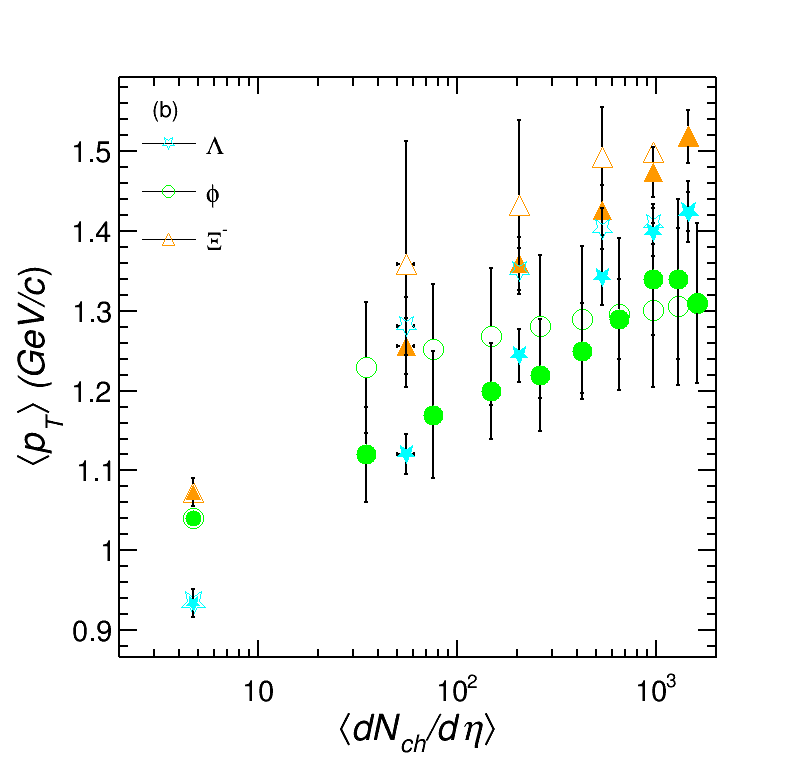}
\caption{$\langle p_T\rangle$ as a function of $\langle \frac{dN_{ch}}{d\eta}\rangle $ 
in Pb-Pb collisions at 
$\sqrt{s_{NN}}$=2.76~TeV:
a) $\pi^+$, K$^+$ and p \cite{Al3};
b) $\phi$ \cite{Al14}, $\Lambda$ and $\Xi^-$. Open symbols - estimated using Eq.(3) 
and full symbols - experimental values (see the text).}
\label{fig-4}
\end{figure} 
The values for $\Lambda$ and $\Xi$ were obtained via interpolation using the values at $\sqrt{s}$=900 GeV and 7 TeV
\cite{Al16, CMS}. The values for the $\phi$ meson were taken from \cite{Al20}.

The  $\langle p_T\rangle _i^{cen}$ 
estimated using Eq.(3) are presented in 
Fig.4 by open symbols for light flavor hadrons as a function of 
$\langle \frac{dN_{ch}}{d\eta}\rangle$.  
$\langle p_T\rangle $ values for $\Lambda$ and $\Xi$ for different centralities
were obtained by fitting  the $p_T$ spectra \cite{Al1, Al2} with the expression from \cite{Bil} and extrapolating the measured spectra in the 
unmeasured regions using the fit results.
The experimental results 
are presented by full symbols. As in the case of normalised $p_T$ distributions ratios, the $\langle \frac{dN_{ch}}{d\eta}\rangle $ dependence 
of $\langle p_T\rangle _i^{cen}$ is qualitatively reproduced by a core-corona interplay. The quantitative difference, larger than the one for the yields (Fig.2),
as it was already mentioned, could be due to the fireball shape dependence of the $\frac{d^2N}{dydp_T}$ distribution as a function of
$\langle \frac{dN_{ch}}{d\eta}\rangle $ (centrality).
Based on these results and the following approximation for the transverse energy: 
\begin{equation}
\frac{d{E_T}}{dy}\approx 3\left(\frac{d{E_T}}{dy}\right)_{\pi^{+}}+ 
4\left(\frac{d{E_T}}{dy}\right)_{{K^{+}},p,\Xi^{-}}+
2\left(\frac{d{E_T}}{dy}\right)_{\Lambda,\Omega^{-}}
\end{equation} 
\noindent
where $\frac{d{E_T}}{dy}=\langle m_T\rangle \frac{dN}{dy}$ and $\langle m_T\rangle =\sqrt{\langle {p_T}\rangle ^2+m^2}$, 
the influence of the interplay between core and corona contributions on the 
$\langle \frac{dE_T}{dy}\rangle /\langle \frac{N_{part}}{2}\rangle $
can be also estimated.
\begin{figure}[h]
\centering{\mbox{\epsfig{file=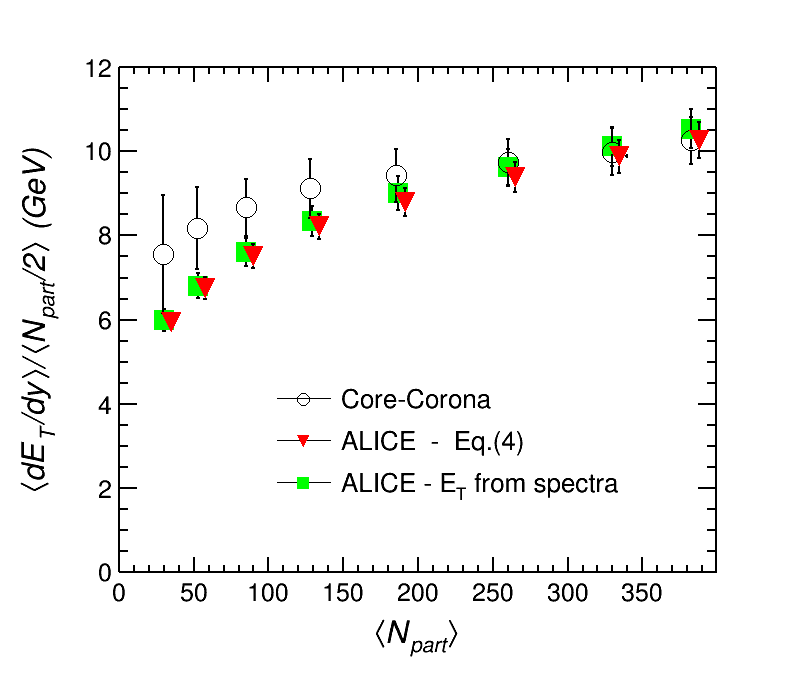, width=.47\textwidth}}}
\caption{$\langle \frac{dE_T}{dy}\rangle$/$\langle \frac{N_{part}}{2}\rangle$ as a function of $\langle N_{part}\rangle$ 
for Pb-Pb collisions at $\sqrt{s_{NN}}$=2.76 TeV:
by using Eq.(4) and the core-corona approach for $\frac{dN}{dy}$ and
$\langle m_T\rangle $ - open circles;
the results obtained using the experimental values (see the text) - red triangles; 
$\langle \frac{dE_T}{dy}\rangle$ values from Ref. \cite{Al15} - green squares. To be visible, the red triangles are shifted by 5 units in 
$\langle N_{part}\rangle$.}
\label{fig-5}
\end{figure} 
As could be noticed, Eq.(4) does not include the contribution of $\Sigma^+$,
$\Sigma^-$ and the corresponding anti-baryons. The estimate of their contribution to $\langle \frac{dE_T}{dy}\rangle$, based on different assumptions and model predictions \cite{Al15,Al21}, is $\sim2.75\%$, well within the experimental errors and model dependent predictions. 
As far as this does not change the conclusion on the $\langle N_{part}\rangle$ dependence of
$\langle \frac{dE_T}{dy}\rangle$ in the 
core-corona approach and the experimental $\Sigma$ spectra are not yet available, we decided to not include the contribution of $\Sigma$ baryons in the present study.     
The results are presented in Fig.5 by open circles. The $\frac{dE_T}{dy}$ values estimated within the core-corona approach 
follow the trend obtained using in Eq.(4) 
the experimental values - red triangles. This estimate is in a very good agreement with the $\langle \frac{dE_T}{dy}\rangle $ values from Ref. \cite{Al15} - green squares. 
\begin{figure}[h]
\centering{\mbox{\epsfig{file=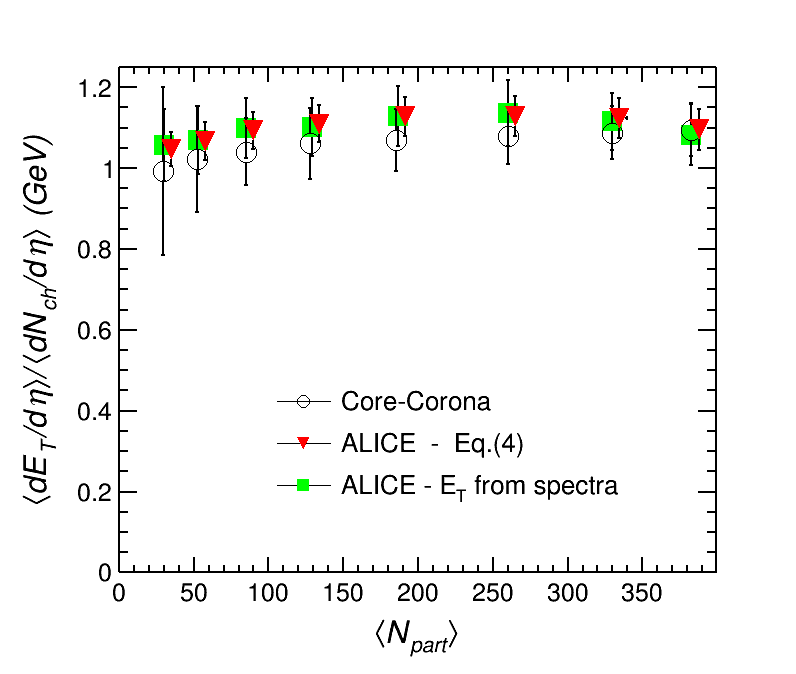, width=.47\textwidth}}}
\caption{$\langle \frac{dE_T}{d\eta}\rangle $/$\langle \frac{dN_{ch}}{d\eta}\rangle $ as a function of $\langle N_{part}\rangle $ for Pb-Pb collisions at $\sqrt{s_{NN}}$=2.76 TeV. The core-corona estimates are represented by open symbols, the ratios obtained using in Eq.(4) the experimental values by red triangles and the ratios based on the values from Ref. \cite{Al15} by green 
squares. To be visible, the red triangles are shifted by 5 units in $\langle N_{part}\rangle$.}
\label{fig-6}
\end{figure} 
In Fig.6 the ratio $\langle \frac{dE_T}{d\eta}\rangle$/$\langle \frac{dN_{ch}}{d\eta}\rangle $  as a function of $\langle N_{part}\rangle$ is represented. The ratio obtained based on the core-corona estimate - open symbols agrees 
with the experimental values within the error bars. This shows that in such a representation the $\langle N_{part}\rangle$ dependence of $\langle \frac{dE_T}{d\eta}\rangle $ and $\langle \frac{dN_{ch}}{d\eta}\rangle $ has a similar trend
due to the interplay between core and corona contributions, therefore it cancels in the ratio.

In conclusion, the centrality dependence of the 
light flavor hadrons yields at mid-rapidity
is well reproduced by a simple core-corona assumption where the corona represents the 
percentage of the wounded nucleons which scatter only once and the core represents the rest of wounded nucleons at a given collision centrality.
The difference between the experimental 
ratio of normalised $p_T$ distributions to the average charged particle densities for Pb-Pb and minimum bias pp collisions at the same energy, the average transverse momenta and the transverse energy at mid-rapidity, as a function of centrality and the estimates 
using the core-corona approach 
can be attributed to the core shape dependence of different phenomena, like flow and suppression,
which depends on the collision geometry and is not included in the present approach. 
The broad distribution of $N_{part}$ for a given experimentally selected centrality 
is also not considered in the present estimates. 
The results show that the corona contribution plays an important role also 
 at energies available at the Large Hadron Collider and it has to be separated in order to evidence the centrality dependence of different observables related to the 
 core properties and dynamics.  

\begin{acknowledgments}
This work was carried out within the 44/05.10.2011 and 2/03.01.2012 projects sponsored by Ministry of Research and Inovation via CNCSI and IFA coordinating agencies. 
\end{acknowledgments}

\bibliographystyle{unsrt}

\end{document}